# Investing in crypto: speculative bubbles and cyclic stochastic price pumps

Misha Perepelitsa

Department of Mathematics

University of Houston

**Abstract**

The problem of investing into a cryptocurrency market requires good understanding of the processes that regulate the price of the currency. In this paper we offer a view of a cryptocurrency market as an environment for realization of a self-organized speculative scheme that results in a formation of a characteristic price bubble as a transient phenomenon. We use microscale, agent-based models to simulate the system behavior and derive macroscale ODE models to estimate such parameters as the return rate and the market value of investments. We provide the formula for the total risk of the system as a sum of two independent components, one being characteristic of the price bubble and the other of the investor behavior.

## 1. Introduction

Price bubbles are ubiquitous in trading of commodities. They range in their characteristics and motive powers that bring them about such as herding behavior, trend chasing, technical analysis or unfulfilled expectations. Cryptocurrencies are the novel type commodities that combine some specific properties such as absence of the "fundamental'" valuation, virtually infinite divisibility, and global and cheap accessibility through Internet. Not surprisingly, price bubbles that are bound to appear in cryptocurrency markets will have some unique properties too. Perepelitsa and Timofeyev (2019, 2021) considered a scenario of a bubble formation that utilizes the above-mentioned properties of cryptocurrencies. In that model, called asynchronous stochastic price pump (ASPP), agents trade a commodity (stock) based on their expected "stock-to-cash" ratios which are updated after a trade. An agent stock-to-cash ratio is increased if the agent sells stock and is decreased otherwise by some factors that we descriptively call the greed and fear factors (levels). Trading sessions take place regularly and anonymously (as on Internet), i.e., between randomly selected traders whose preferences are kept private. As one can expect, if the greed factor exceeds the fear factor, the price of the commodity will grow. The price series generated by an ASPP has returns following log-normal distribution and the mean return can be estimated in terms of the greed and fear factors and other parameters of ASPP. The model shows self-organization expressed in collectively produced positive price dynamics, when each agent follows an adaptive strategy perusing self-interest. An ASPP system is conservative with no in- or out-flow of money. In the trading process smaller and smaller amounts of commodity are bought and sold at increasing price, which makes infinite divisibility of a cryptocurrency well suited for the purpose. The trading process could go ad infinitum but it will crash.

The difficulty of knowing when the bubble crashes is an absence of any sort of fundamental value to measure price deviations. There is however an inherited risk of a crash that in Perepelitsa and Timofeyev (2019) was identified with the process of re-distribution of cash among traders during a lifespan of ASPP in which a sizable group of traders run out of cash and become heavily invested in stock, while the other group accumulate most cash. Many trading processes in random environments have this property (Dragulescu and Yakovenko 2000, Basi et al. 2009, Cordier et al. 2005). Such redistribution creates a conflict between an individual inside the low-cash-group and the collective behavior: increasing price is a collective effort, but at some point, the individual would like to sell her stocks, while she prefers to keep buying. Such an unstable state may lead to a sudden sellout and a subsequent crash. This is endogenously generated systematic risk (formula (8) on page 5).

After a crash, if the greed and fear factors do not change much, ASPP will continue pumping price up and generate more risk until the next crash. We would like to associate the ASPP model with the early stage of Bitcoin market, but ASPP being a conservative model, is probably is an underestimation because the capitalization of Bitcoin steadily increased over the years. In recent years the amount of investment to Bitcoin and other cryptocurrencies has surged, marking a new "investment phase" of the cryptocurrencies. We would like to model it as a long-term investment into a price bubble produced by ASPP. The questions then arise naturally, how the dynamics of an ASPP changes during the investment phase, in particular, what the rate of return and the systematic risk are, and how to define the total risk of the new system and what predictions can be drawn from this model. These are questions we address in the present paper.

In section 2 we analyze the behavior of investment ASPP model using multi-agent simulations. An informative way to look at the dynamics is to map it in the plane of log of price vs. hazard rate, as for example in Figure 4, which gives a typical description of the investment cycle, showing its main stages. In the first stage, ASPP works by itself generating high return rate and accompanying systematic risk. It is followed by the initial investment stage with positive inflow of capital. This further enhances the return rate and reduces the systematic risk of the ASPP giving the price bubble a new life. At the third stage, as investors start to make use of high returns and withdraw capital the system transitions into regime of high price oscillations with growing systematic risk. At this stage, alongside with the systematic risk, there appears the risk from the investor side of the model, as the oscillating price underperforms the initial expectations. Due to this enhanced risk the system is likely to crash at some point during this last stage, finishing the investment cycle. Interestingly, by the end of the cycle there is a positive net inflow of cash into ASPP component, although it is not distributed equally among traders.

As the mathematical analysis of a non-linear multi-agent, stochastic system is difficult a task we seek a lower dimensional model that captures main features of the dynamics of multi-agent system. Using deterministic ODE model of a classical Ponzi scheme, and the trading mechanism by which the price is set in ASPP model we derive a non-linear ODE model, called speculative Ponzi scheme that we use to estimate the value of the investment in the system and the market return rate at time $t$ in section 2. It turns out that speculative Ponzi scheme is quite accurate approximation of the transient behavior of multi-agent ASPP system.

The main feature of the transient behavior is a characteristic price bubble. The growth of the bubble is driven by positive influx of money that reinforces itself. During that time the returns accelerate. A reasonable assumption is an exponential growth of investments. This growth phase lasts for initial maturity period and end when the investors start to withdraw a proportion of realized profit, see Figure 3 on page 7. As the rate of withdrawals exceeds the rate of investments the price drops. Note, that the burst of the bubble occurs even assuming that the investments are still coming at the same rate as before the crash. In reality they necessarily drop enhancing the negative price dynamics. In any case, even with mildly growing investments, after the crash the schemes enters a stationary regime with zero returns, which can be considered the end of the speculative scheme. The role of the blockchain in this process is to provide the environment (platform) for realization of the bubble.

The speculative Ponzi component of the investment system carries its own systematic risk, which is however is easier to identify as the gap between investors expected and the actual, realized return, which is typically less than expected return. The total risk (hazard rate) is defined by formula (12) as a sum of the risks of the ASPP and the speculative Ponzi components.

## 1.1 Literature overview

Bubbles in financial markets is well researched area. Large price fluctuations in stock prices can be explained by the presence of noisy traders that imitate other traders and look at the mar- ket trends (Bak et al. 1997, Levy et al. 1994, 1995, 1997, Lux 1995, 1998, Lux and Marchesi 1999, 2000). Price fluctuations can arise in the process of "genetic" evolution of trading strategies, when traders constantly adopt and modify existing ones, selecting better performing strategies, as in the model by Arthur et al. (1997). Price bubbles may arise from rational expectations of traders and be consistent with efficient market hypothesis. This type of price dynamics was introduced by Mandelbrot (1966) and extended by Blanchard (1979), Blanchard and Watson (1982). Within this framework, Johansen, Ledoit & Sornette (1999, 2000), Johansen and Sornette (1999), Sornette (2003) constructed realistic non-linear models of price series that are characterized by price bubbles and crashes.

Agent based modeling is by now a well-developed and widely used methodology. The review of agent-based modeling in economics and finance can be found in Chakraborti et al. (2011), Chen et al. (2012), Iori and Porter (2014). There are few papers that apply the agent-based modeling to the cryptocurrencies trading. On this topic we would like to cite works Cocco, Tonelli & Marchesi (2019, 2019b) who analyze the price dynamics generated by the traders using the genetic algorithms, technical analysis, or random strategies.

Redistribution of wealth among traders in random trading processes was discussed by Dragulescu and Yakovenko (2000), Basi et al. (2009), Cordier et al. (2005). Mathematical models for Ponzi schemes can be found in Artzrouni (2009). In Perepelitsa (2021) authors studied the effects of heterogeneous distribution of levels of greed and fear in a population of trades on the dynamics of ASPP and performed a preliminary statistical data analysis on estimating these factors from Bitcoin price series.

## 2. Method

### 2.1 ASPP model

ASPP is a model of a speculative bubble in a price series of a commodity that doesn't have an underlying fundamental value and is infinitely divisible and thus, as was argued by Perepelitsa and Timofeyev (2019, 2021) well-suited for the study of price fluctuations of cryptocurrencies. We recall the model and its main properties from those references. In the model, the market consists of a set of $N$ agents, described by their portfolios $(s_i, b_i)$, $i = 1..N$, of dollar values of a stock and cash accounts, and let $k_i$ stand for agent $i$ target stock-to-bond ratio. $P_0$ will denote a current price per share and $P$ the new price determined by agents' demand. Let $\{i_l \mid l = 1..m\}$ be the set of "active" agents, i.e. the ones setting the new price. The set of active traders is determined each trading period by a random draw from the population. If $x_{i_1}$ is the dollar amount that agent $i_1$ wants to invest in stock, then it is determined from the condition:

$$\frac{\frac{P}{P_0} s_{i_l} + x_{i_l}}{b_{i_l} - x_{i_l}} = k_{i_l}.$$

( 1 )

The demand-supply balance is a simple market clearance condition:

$$\sum_{l=1}^{m} x_{i_l} - x_{in} = 0,$$

where $x_{in}$ is the amount of the exogenous investment (positive or negative) into the system per trading period. Using equation (1) in the last equation we can solve it for $P$:

$$\frac{P}{P_0} = \left( x_{in} + \sum_{l=1}^{m} \frac{k_{i_l} b_{i_l}}{1 + k_{i_l}} \right) \left( \sum_{l=1}^{m} \frac{s_{i_l}}{1 + k_{i_l}} \right)^{-1}.$$

( 2 )

Once the price is set, agents move corresponding amounts between cash and stock accounts, re-balancing their portfolios. At the next step, the active agents update their expected stock-to-bond ratios from current value $k_{i_l}$ to a new value $\hat{k}_{i_l}$ using the agents specific greed ($\alpha_{i_l} \geq 1$) or fear ($\beta_{i_l} \geq 1$) factors, depending on whether the agent sold or bought stocks:

$$\hat{k}_{i_1} = \begin{cases} \alpha_{i_l} k_{i_l}, & \frac{P s_{i_l}}{P_0 b_{i_l}} > k_{i_l}, \\ k_{i_l}, & \frac{P s_{i_l}}{P_0 b_{i_l}} = k_{i_l}, \\ \frac{k_{i_l}}{\beta_{i_l}}, & \frac{P s_{i_l}}{P_0 b_{i_l}} < k_{i_l}. \end{cases}$$

( 3 )

The trading session is repeated the following trading periods with new, randomly selected sets of active agents. The behavior of agents in (3) can be described as *adaptive*. If an agent portfolio under-performs, i.e., doesn't meet the expected stock-to-cash, ratio, the agent will decrease the investment in stock, and visa-versa.

Below we recall properties of the ASPP model with zero external investment, $x_{in} = 0$, obtained by Perepelitsa and Timofeyev (2019, 2021). The model has a stationary stable state when all $\alpha_i = \beta_i = 1$, i.e., when traders do not change their stock-to-cash ratios after trading. In such situation, if the ratios are balanced, no trading takes place the price remains constant. This simple regime has an interesting property of {\it mimicking} the fluctuations in the cash reserves. For example, if the "cash" part of the investment represents investment in some other commodity that grows (drops) at a certain rate then ASPP model will generate a price series with a positively correlated price series.

Our main interest is the case when the update rule (3) changes the agents investment ratios, that is, when $\alpha_i, \beta_i > 1$. The model also has a stationary state when all investment ratios are balanced and price is constant, but this stationary state is unstable, with tiny perturbations leading to divergent dynamics in the price series.

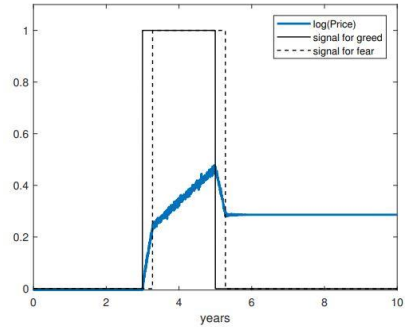

Figure 1: Propagation of greed and fear signals through ASPP. The level of greed changes as $\boldsymbol{\alpha_i = 1 + \widetilde{\alpha}}$ · (signal), and fear $\boldsymbol{\beta_i = 1 + \widetilde{\beta}}$ · (signal). In a greed dominating fear regime, $\widetilde{\alpha} > \widetilde{\beta}$, the price increases and the system reaches new stable steady state. Plot shows one path simulation of an ASPP with 500 agents and 125 activagents per daily trade.

Figure 1 shows an example of the price series generated by an ASPP when a signal of greed, $\alpha_i > 1$, followed by a fear signal $\beta_i > 1$, propagate through the population of the traders. In this example greed dominates fear, $\alpha_i/\beta_i > 1$ and the price increases to a higher value. With the levels of $\alpha_i, \beta_i$ dropping to 1, the system returns to a stable steady state. Thus, trading activity in the system can be intermittent with periods of non-unit fear and greed levels, followed by unit levels, and any level of price can be realized as stable regime in ASPP. In this paper we consider the situation when the system is always "turned on" by signals of greed and fear. In this case, $\alpha_i$ and $\beta_i$ do not depend on time. This is a likely assumption due to incessant stream of news in media about to cryptocurrencies that feeds into the informational input of agents or even due to feedback of the price fluctuations to the psychological state of agents.

For values of $\alpha_i = \alpha, \beta_i = \beta$, close to 1, in Perepelitsa and Timofeyev (2019), we showed that the logarithm of the return, R(t), of a price series of an ASPP, in a long run, is distributed according to a normal distribution

$$\ln\big(R(t)\big) \in \mathcal{N}(\ln r^*, \ \sigma^*),$$

$$(4)$$

with the formulas the rate of return $r^*$ (geometric mean) is expressed by formula

$$r^* = \left(\frac{\alpha}{\beta}\right)^{\frac{1}{\gamma}}, \quad \sigma^* = c_0(\alpha\beta - 1),$$

$$(5)$$

where $\gamma = 2N/m$. In this formula, $N$ is the number of traders, $m$ in the number "active traders", that is, the traders participating in buying and selling per day. $c_0$ is a positive constant. For a reference, we note that in Perepelitsa and Timofeyev (2019, 2020), the inverse of $\beta$ was used instead in formulas for $r^*$ and $\sigma^*$. The above model applies to the situation when all agents in the system use the same multipliers $\alpha$ and $\beta$ to change the investment ratio. The effect of varying parameters $\alpha$ and $\beta$ across the population on the performance of ASPPs was considered by Perepelitsa (2021) where it was shown that formulas sill applying when values $\{\alpha_i\}$ and $\{\beta_i\}$ perfectly correlated (coefficient of correlation equals 1), but the rate of return $r$ decreases with the decreasing correlation. In the ASPP models that we use for illustration in this paper we consider heterogeneous distribution of $\{\alpha_i\}$ and $\{\beta_i\}$ with correlation coefficient $\rho = 0.95$.

Being an agent-based model, ASPP contains complete information about the state of the system, including the information about the distribution of investments of agents. Following Perepelitsa and Timofeyev (2021), we use that information to quantify the systematic risk of a crash of a price bubble.

By systematic risk we mean the risk generated endogenously by ASPP. Adaptive trading (3) leads to dispersion of the distribution of cash (and stocks) among agents. A situation may arise when a significant group of agents end up having low cash reserves and high stock investment. This creates a conflict between an individual inside this group and the collective behavior: increasing price is a collective effort, but at some point, the individual would like to sell his/her stocks while he/she prefers others to keep buying. Such an unstable state may lead to a sudden sellout and a subsequent crash. To quantify the risk, we let $p(s,t)$ be the (probability) distribution of agents according to the cash reserves. Function

$$h(t) = \int_0^\infty e^{-s^2/\gamma_1} p(s,t)\,ds$$

( 6 )

measures the amount of concentration of agents with low cash. From this definition, it follows that $h(t) \in (0,1]$, with $h(t) = 1$, meaning the all agents have zero cash. The hazard rate for the crash can be defined as a monotonly increasing scale function from $[0,1] \rightarrow [0,+\infty)$. In the examples in this paper we will use function

$$H_a(t) = \frac{\gamma_2\sqrt{h(t)}}{1 - \sqrt{h(t)}}.$$

( 7 )

In the above formulas (6) and (7), $\gamma_1, \gamma_2$ are positive constants.

### 2.2 Benchmark models for Ponzi schemes

*2.2.1 Classica Ponzi schemes* First, we recall a mathematical model of a classical Ponzi scheme, from example Artzrouni (2009). In such a model, investors are guaranteed a certain rate of return $r_p$ (promised rate) after a maturity period $t_m$. Other scheme parameters include: the nominal rate of return $r_n$, which can be, for example, the current market rate, and $r_w$ - rate of withdrawal (target rate). Suppose that money is invested at the rate $r(t)$ dollars per year, called the investment schedule. Denote by $S(t)$ - total money in the Ponzi scheme at time t, and $R(t)$ – value of invested funds available for withdrawal, as anticipated by the investors. Dynamics of $S(t)$ is described by an ODEs:

$$\frac{dS}{dt} = r_n S(t) + r(t) - r_w R(t), \qquad S(0) = S_0,$$

$$\frac{dR}{dt} = r_p R - r_w R + e^{r_p t_m} r(t - t_m), \qquad R(0) = 0.$$

( 8 )

Here we assume that $r(t)$ is zero when $t < 0$. The solution to the second equation yields function R(t) which is zero for $t \in [0, t_m]$, and for $t > t_m$:

$$R(t) = \int_0^{t-t_m} e^{(r_p - r_w)(t - t_m - s) + r_p t_m} r(s)\,ds.$$

( 9 )

Given $R(t)$ one can easily solve the first equation for $S(t)$. Assuming that $r_p > r_n$ (otherwise the scheme is not attractive for investors), the typical question is when the scheme runs out of money ($S(t) = 0$), given the initial money in the system $S_0$ and investment schedule $r(t)$.

Figure (2) shows life stories of several Ponzi schemes with different investment schedules. For the scheme with the constant rate $r(t) = const.$, the capitalization of the scheme grows well beyond the $t = t_m$ the initial maturity period. This may act as positive feedback on the rate of investment, so one might consider growing investment rates. Figure (2) shows dynamics of Ponzi schemes with two such accelerating flow of investments: linear and exponential schedules.

Note that investment schedule $r(t)$ is the principle parameter that cannot be controlled directly by the originator of a Ponzi scheme. It is a key element of the model that the originator may try to influence by offering an unreasonably high promised rate $r_p$. Surprisingly (or not) this strategy has been working if we review the known historical instances of Ponzi schemes. The lifespan of the scheme varies for different investment schedules, but for exponential schedule $r(t) = e^{\alpha t}$, there is a critical value $\alpha_c$ above which ($\alpha > \alpha_c$) the scheme is viable for all times $t > 0$. The latter scenario can be realized, for example, if the scheme operates in timescales of several generations of investors, naturally growing at the exponential rate, as implemented in a social security model. Consider, for example, the case $r_n = 0$ (flat market rate), and withdrawal rate $r_w$ equal to the promised rate $r_p$, so that after the maturity period investors withdraw a fixed sum per unit of time. In this setup, one can compute

$$\frac{dS}{dt} = e^{\alpha t} - e^{r_p t_m} r_p \alpha^{-1} \left( e^{\alpha(t - t_m)} - 1 \right).$$

It is easy to see that $\alpha_c = r_p$ is the critical value. The exponential investment schedule $r = c e^{\alpha t}$, changes its behavior at the critical value $\alpha = \gamma$, after which the scheme becomes viable for all times $t > 0$. More relevant, in the context of Bitcoin, is the possibility that global access to cryptocurrencies market through Internet and universal contentedness through global information networks create the critical (self-sustained) growth of investments.

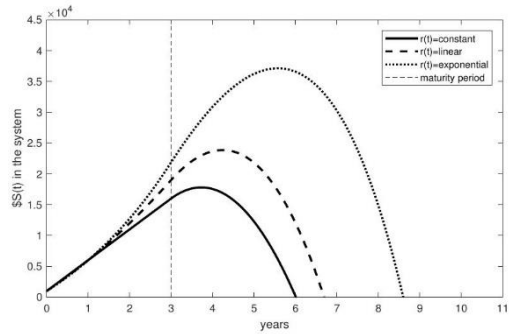

Figure 2 Three classical Ponzi schemes. The plot shows capitalization of Ponzi schemes for three investment schedules: $r(t)$=constant, $r(t)$=linear, $r(t) = c_1 e^{0.3t}$ normalized to have the same investment during first year. Rates: $r_n = 0, r_p = r_t = 0.41$. Maturity period $t_m = 3$ years.

### 2.2.2 A self-organized Ponzi scheme

By a speculative Ponzi scheme, we mean an investment mechanism in which the nominal (market) rate is determined by the collective behavior of investors. The key assumption the following approximation for the nominal rate, that can be obtained from equation (2) assuming that an ASPP is indeed the engine that sets the price. According to this formula we can take the nominal (market) rate to be proportional to the influx of money into the scheme plus a rate $r^*(t)$ generated by an ASPP:

$$r_n(t) = c_0 \left( r(t) - r_w R(t) \right) + r^*(t), \quad c_0 = const.$$

$$( 10 )$$

To simplify the analysis, we will assume that $r^*(t) = 0$ in (10), so that $r_n(t)$ completely determined by the balance of in- and out-flow of money in to the system. Since in this model no promises are given, the promised return rate must be set equal to the nominal rate, given by (10). The resultant differential equations are now nonlinear:

$$\frac{dS}{dt} = \left( r(t) - r_w R(t) \right)(c_0 S + 1), \quad S(t = 0) = S_0,$$

$$\frac{dR}{dt} = (r_n(t) - r_w)R + r(t - t_m)e^{\int_{t-t_m}^{t} r_n(s)\,ds}, \quad R(t=0) = 0,$$

$$r_n(t) = c_0\big(r(t) - R(t)\big), \quad t > 0.$$

Figures (3) illustrates the dynamics of the speculative Ponzi schemes for constant, linear, and exponential investment schedules. The characteristic feature of this dynamics is a formation and a burst of a transient bubble.

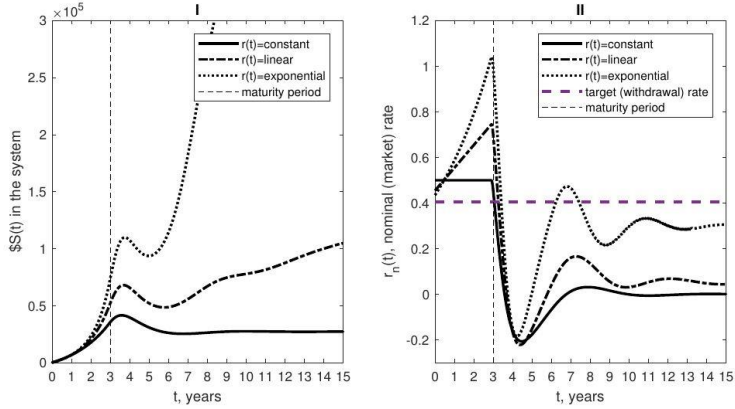

Figure 3: Three self-organized, speculative Ponzi schemes. Left plot I: value S(t) of the investment at time t. Right plot II: nominal (market) rate $r_n(t)$ at time t. Investment schedules are $r(t)$=constant, $r(t)$=linear, $r(t) = c_1 e^{0.1t}$ normalized to have the same investment during first year. Rates and maturity period as in Figure 2.

In comparison to the classical Ponzi case discussed above, speculative schemes do not run out of money, but they have inherited risk of a crash (see below). Capitalization of such schemes grow much faster during the initial maturity phase and then experience a crash $t > t_m$, reducing the value of the initial investments. Figure (3) shows a drop in nominal return rate to negative values for several year after $t_m$. This period can be taken as the end the life of the scheme. However, if the crash didn't scare investors (investment schedules doesn't change), the schemes for all three investment schedules remain viable for all times and after few oscillations reaches a steady state with rate of the nominal return $r_n$ below the target (withdrawal) rate $r_t$. The steady state return for constant and linear investment schedules are equal to zero and eventually the investors will lose interest in it. For exponential investment schedules the nominal (market) rate approaches some positive limiting value: $\lim r_n(t) = r_s > 0$. In either case, the investment of a \$1(in a stationary regime) results in the positive return of \$$r_w/(r_w - r_s)$, over the infinite horizon of the investment, provided that the investments keep coming at the exponential rate, which is not realistic. As in the classical Ponzi scheme there is a critical value $\alpha_c$ for the exponential investment schedules $r(t) = e^{\alpha t}$ when at the steady state the nominal (market) rate matches the target (withdrawal) rate.

This mismatch between actual return $r_n(t)$ and target return $r_w$ generates a risk of a crash when the investors may spontaneously and collectively decide to withdraw from the scheme. This is another component of a total risk, not included in the risk of crash of ASPP given by (7). A possible metric for this new risk is the unrealized profit:

$$H_p(t) = \gamma_3 \int_{t_m}^{\max\{t,t_m\}} e^{r_w - r_n(s)}\,ds,$$

(11)

for some positive $\gamma_3$. With this we obtain the formula for the total systematic risk of ASPP model with investments as

$$H(t) = H_a(t) + H_p(t),$$

(12)

where $H_a$ and $H_p$ given by (7) and (11), respectively.

## 3. Results

We consider the dynamics of ASPPs in log(Price)-hazard plane, by plotting points $\big(\log P(t), H_a(t)\big), t \in [0..T]$,



for several representative cases of ASPPs.

### 3.1 Zero investment regime: conservative ASPP

When the flow of investments is zero, $x_{in} = 0$, ASPP price bubble pumps the price at the rate close to $r^*$ in (5) and the hazard rate $H(t)$ grows monotonically due to unequal re-distribution of cash caused by trading. On Figure 2 this process is represented by a curve OB.

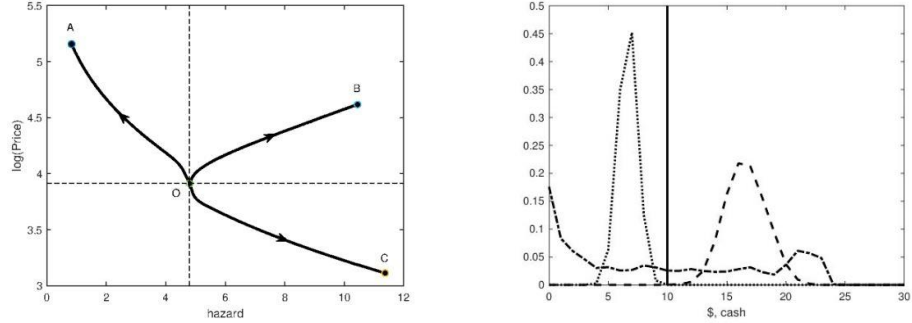

Figure 4: Left: ASPP dynamics in log(Price)-hazard plane. Curves represent the simulated dynamics of ASPP with large positive rate of investments (OA), zero-investment (OB), and large withdrawal rate (OC). Investment into ASPP reduces the systematic risk of a crash. Right: Distribution of cash among agents in ASPP system. At time $t = 0$, all agents have \$10 (represented by a solid line on the graph). Dashed line corresponds to state A, dash-dotted line corresponds to state B, and dotted line to state C.

### 3.2 ASPP in an investment regime

When the flow of investments is positive $x_{in} > 0$, the system continuously transitions (for sufficiently large rate on investments) to a regime shown by curve OA in Figure 2 (right panel).

Here, the investments contribute to higher rates of return and have effect of diminishing the systematic risk expressed by the hazard rate $H(t)$. In this regime there is positive inflow of cash into ASPP system. When the investment is withdrawn from the system, i.e. when $x_{in} < 0$ and sufficiently low, the system continuously transitions to a regime represented by curve 0C. The price goes down and the hazard rate increases. At such process, cash is being removed from ASPP. Figure 4 (right panel) shows the distribution of cash among agents at the final stages A, B and C of processes shown in Figure 4 (left panel).

### 3.3 ASPP in a full investment cycle

Now consider a full investment cycle based on ASPP model. At the initial stage (for the first 3 years) an ASPP runs with zero external investments, producing a price bubble with high return rates, close to $r^*$ from (5). Then, it is followed by an investment phase, when investors put money into ASPP that they keep there during maturity period of $t_m = 3$ years, after which they start to withdraw money at the "expected," target rate $r^*$. We refer to this process as an investment cycle. We denote the flow of money per year into ASPP by function $r(t)$ (investment schedule) and assume that it is a priori given and is not affected by the changing dynamics of an ASPP.

Figures 6 illustrate main characteristics of the ASPP system during an investment cycle. In this example, the investment schedule is chosen to be exponential $r(t) = c_1 e^{0.1t}$, where $c_1$ is selected so that the total investment during the first year matches cash reserve of the ASPP (strong investment regime). More examples, with different investment schedules are provided at the end of this section. Consider the dynamics of the ASPP in the space of log(Price)-hazard rate parameters, Figure 6, right panel. Zero investment initial phase, AB generates a price bubble with high return and is followed by the investment phase. The initial maturity period, BC, lasts from year 3 to 6, and reduces the systematic risk $H_a(t)$, defined in (7). When investors start to withdraw money, the price sharply drops and afterward goes through large scale oscillations, finally stabilizing at the rate below the target rate $r^*$. This stage is represented by curve CD in the phase space. The systematic risk increases during this stage, but also, the risk is amplified due to the fact that investors are not getting the expected rate. This combined risk (see formula (12) below) will result in crash somewhere between stage C and D. It is sketched as a dotted line EF in Figure 6, right panel. Thus, the dynamics of the investment cycle will proceed along the path ABCEF. Figure 6, right panel shows that during the whole process there a positive net inflow of money into ASPP system, implying that after the crash ASPP component of the system is likely to start a new "cycle".

The main characteristics of this process are the following: a) the cycle runs on cash of greedy investors that are enticed by the high returns of the initial phase of the cycle of ASPP; b) the cycle results in positive net transfer of



cash from investors to ASPP component; c) the cycle is robust: it can last for prolonged periods of time and is repeatable. The economic significance: the cycle leads to re-distribution of cash among investors and ASPP agents, by concentrating cash among a (small) group of participants. Psychological significance: the cycle consists in generating among the participants the excitement of anticipation of both, high payoffs and an imminent but unpredictability occurring crash.

Figure 7 illustrates an investment cycle of ASPP with linear, $r(t) = c_3 t + c_4$, investment schedule, where parameters are chosen so that the investment during the first year matches cash reserve in ASPP. In following section, we show the characteristic oscillations of the parameters in the agent-based ASPP model can be described by an ODE system based on a simple model speculative Ponzi scheme.

*3.4 Comparison of agent-based ASPP model with the speculative Ponzi scheme model*

To see how well the speculative Ponzi scheme approximates the agent-based investment cycle of ASPP, we consider example of ASPP with investment schedule $r(t) = c_1 e^{0.1t}$, discussed above. For the speculative Ponzi scheme, we take the same schedule $r(t)$ and target rate $r_w = r^*$. There is only one scalar extra parameter, $c_0$ in (10), which is not specified by the ASPP model. It is selected to produce best fit to the data. Figure 8 shows the plot of the value of investment S(t) and nominal(market) rate $r_n(t)$ as functions of time in agent-based ASPP and the speculative Ponzi scheme. The plots show the good qualitative fit provided by the Ponzi scheme.

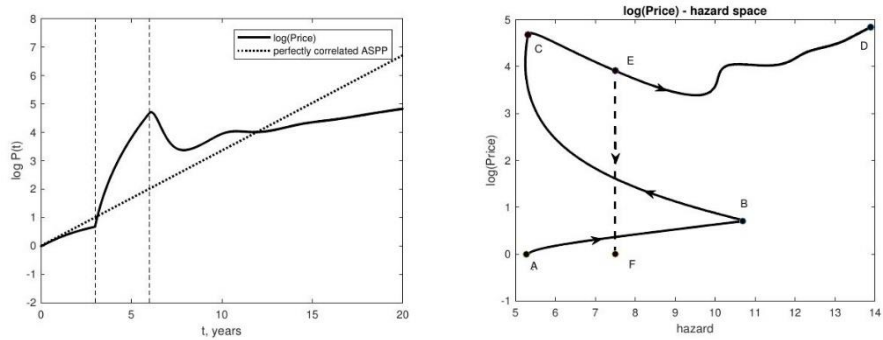

Figure 5: Left: ASPP dynamics: log(Price) time series. The path is an average over 1000 realizations of an ASPP model with exponential investment schedule $r(t) = c_1 e^{0.1t}$ over 20 years, see Appendix for more details. Dotted line is the constant return line, given by (5), corresponding to perfectly correlated, zero-investment ASPP. Right: ASPP dynamics: log(Price)-hazard plane. Zero investment phase AB is followed by initial maturity period BC when the investments flow in at the rate $r(t)$ per year. After the system reaches point C, investments continue to flow but also withdrawn at the rate $r^*$. Segment EF shows the transition of the system after a (probable) crash. Dinamics of investment ASPP proceeds along cylce ABCEF.

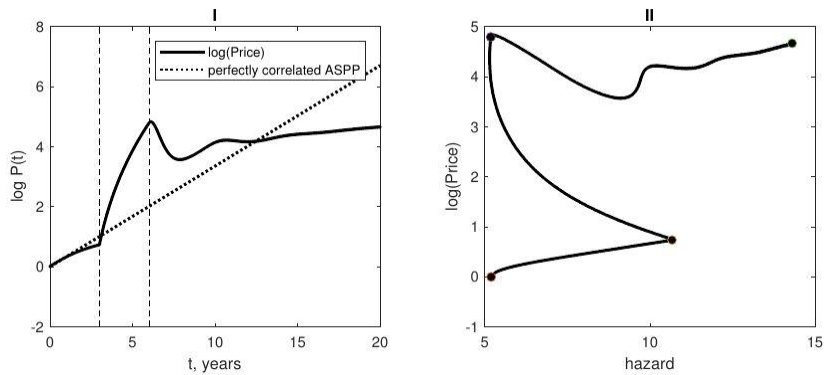

Figure 6: ASPP dynamics: linear investment schedule $r(t) = c_3 t + c_4$. The level of investment in the first year matches initial cash reserve of ASPP.



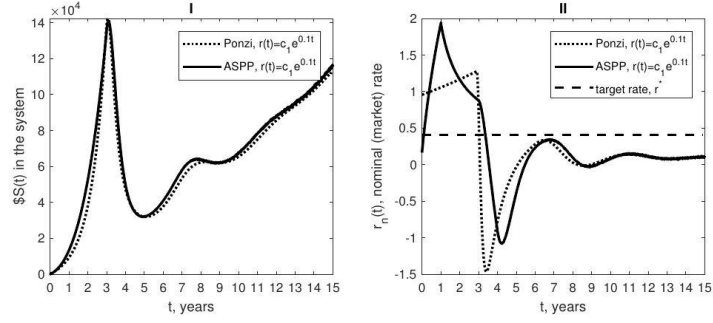

Figure 7: Speculative Ponzi scheme vs. Investment ASPP. Both models have exponential schedule $r(t) = c_1 e^{0.1t}$. Left plot I: value $S(t)$ of the investment at time $t$. Right plot II: nominal (market) rate $r_n(t)$ at time $t$.

## 4. Discussion

In this paper we considered a model of speculative trading that has two superimposed levels of self-organized patterns of speculative trading. One is the adaptive ASPP trading model which, due to specifics of the trading commodity (no fundamental value and infinity divisibility), can generate price bubbles that are stable, in a sense that all risk of failure comes from the endogenous systematic risk. We relate this model to the initial stage of Bitcoin life. Building on the ASPP characteristics (high returns, etc.), another level of self-organization comes from the behavior of investors who start to consider ASPP for investment on a larger time scales than adaptive trading in ASPP, possible lasting for several years.

One effect of this is the (temporal) reduction of the systematic risk of ASPP, that prolongs the life of the bubble, and, in general, non-trivially changes the dynamics of the process. The investment component has its own contribution to the total risk that we estimated in (12), so that at later times the risk of ASPP gets amplified. Moreover, we showed that a simple speculative Ponzi scheme gives a good prediction of the long-time behavior. There appears to be another level of complexity of organization in the current stage of cryptocurrency trading, when several cryptocurrencies are traded jointly a single market. This situation can be models by connecting several ASPP of the type consider in this paper. It will be considered in our future work. Finally let us mention several interesting technical extensions that can be added to the models in this paper. One is a more realistic ASPP model when both $\alpha_i$ and $\beta_i$ are proportional to the amount $x_i$ that agent $i$ trades per period We call it strong adaptation hypothesis as opposed to the weak adaptation used in this paper. The other is that there are several natural feedback loops that can be added to the model. For example, price dynamics can force the change the levels of greed and fear among the ASPP traders, and investment schedule $r(t)$ can be affected the dynamics of trading.

In future work, the models considered in this paper will be evaluated using available data from cryptocurrencies markets. In particular, it will be interesting to determine if investments in Bitcoin operation in "self-sustained" critical regime of Ponzi schemes. That is when the expected return is sustained by accelerating growth of investments, as discussed in section 2.

## 5. Appendix

In the examples of this paper we used ASPP with total of N=500 investors, and m=125 randomly chosen agents to participate in a daily trade (360 trades per year). The values of greed and fear factors for agents are randomly sampled from a jointly Normal distribution with the mean of $\ln \alpha = 1.12$, the mean of $\ln \beta = 1.11$, with equal variances: $Var(\ln \alpha) = Var(\ln \beta) = 12 \cdot 10^{-4}$, and the levels of the correlation coefficient $corr = 0.95$. In this setting, return $r^* = 1.15$. The values of $(\ln \alpha, \ln \beta)$ are restricted to the model relevant range of $\{\ln \alpha \geq 0, \ln \beta \geq 0\}$, and the specific choice of variance guaranties that all sample points within 3 standard deviations from the mean are located in this range. All agents start with \$10 in cash initially, with target ratios $k_i = 1$, $i = 1..N$, and starting price $P_0 = 1$. Agent $i$ has \$$(10k_i + \eta_i)$, where $\{\eta_i\}$ are i.i.d. uniform random variables with values in $[0,0.1]$. Zero investment period is the first three years, and maturity period $t_m = 3$. The hazard rate is determined by $\gamma_1 = 70$ and $\gamma_2 = 5$. We performed 1000 path simulations of ASPP with investments for daily trades for the duration of 20 years, for different choices of investment schedules $r(t)$.

**References**

Arthur WB, Holland JH, LeBaron B, Palmer RG, Tayler P (1997). *Asset Pricing under Endogenous Expectations in an Artificial Stock Market*, in W.B. Arthur, S. Durlauf and D. Lane (Eds.), The Economy as an Evolving Complex System II. Addison-Wesley.

Artzrouni M (2009). *The mathematics of Ponzi schemes,* Math. Soc. Sci., 58:190-201.

Bak P, Paczuski M, Shubik M (1997). *Price variations in a stock market with many agents*, Physica A 246 (3-4), 430-453.

Basi M, Spiga G, Toscani G (2009). *Kinetic models of conservative economies with wealth redistribution*, Comm Math Sci, 7:901-916.

Blanchard OJ (1979). *Speculative bubbles, crashes and rational expectations,* Economics Letters, 3, 387-389.

Blanchard OJ, Watson MW (1982). *Bubbles, rational expectations and speculative markets,* in Crisis in Economic and Financial Structure: Bubbles, Bursts and Shocks, P. Wachtel (Eds.), Lexington Books, Lexington, MA.

Chakraborti A, Toke IM, Patriarca M, Abergel F (2011). *Econophysics review: II. Agent-based models,* Quant. Finance, vol. 11, no. 7, 1013-1041.

Chen SH, Chang CL, Du YR (2012). *Agent-based economic models and econometrics,* Knowl. Eng. Rev., vol. 27, pp. 187-219.

Cocco L, Tonelli R, Marchesi M (2019). *An agent-based artificial market model for studying the bitcoin trading,* IEEE Access 7:42908‐42920 DOI 10.1109/ACCESS.2019.2907880

Cocco L, Tonelli R, Marchesi M (2019b). *An agent-based model to analyze the bitcoin mining activity and a comparison with the gold mining industry,* Future Internet 11(1):8 DOI 10.3390/fi11010008.

Cordier S, Pareschi P, and Toscani G (2005). *On a kinetic model for a simple market economy,* Journal of Statistical Physics, 120:253-277.

Dodd N (2017) *The social life of bitcoin. Theory, Culture & Society,* 35 (3) 35-56.

Dragulescu A, Yakovenko VM (2000). *Statistical mechanics of money,* European Physical Journal B, 17(4):723-729.

Feng M, Bai Q, Shan J, Wang X, Chiang R (2015). *The Impacts of Social Media on Bitcoin Performance,* ICIS.

Garcia D, Schweitzer F (2015). *Social signals and algorithmic trading of Bitcoin,* Roy. Soc. Open Sci., vol. 2, no. 9, Art. no. 150288, doi: 10.1098/rsos.150288.

Griffin JM, Shams A (2020). *Is Bitcoin Really Untethered?* J. of Finance, vol. LXXV, no. 4.

Iori G, Porter J (2014). *Agent-based modeling for financial markets,* in The Oxford Handbook of Computational Economics and Finance. S.-H. Chen and M. Kaboudan, Eds. Oxford, U.K.: Oxford Univ. Press.

Johansen A, Sornette D, Ledoit O (1999). *Predicting financial crashes using discrete scale invariance,* J. of Risk 1, 5-32.

Johansen A, Sornette D (1999). *Critical crashes,* Risk, 12(1), 91‐94.

Johansen A, Ledoit O, Sornette D (2000). *Crashes as critical points*, Int. J. of Theoretical and Applied Finance, 3, 219-225.

Levy M, Levy H, Solomon S (1994). *A microscopic model of the stock market: Cycles, booms, and crashes,* Economics Letters, 45:103-111.

Levy M, Levy H, Solomon S (1995) *Microscopic simulation of the stock market: The effect of microscopic diversity,* Journal Physique I, 5:1087-1107.

Levy M, Levy H, Solomon S (1997*). New evidence for the power law distribution of wealth,* Physica A, 242:90-94

Levy M, Levy H, Solomon S (2000). *Microscopic Simulation of Financial Markets: From Investor Behavior to Market Phenomena,* Academic Press.

Liu Y, Tsyvinksi A (2018). *Risks and Returns of Cryptocurrency,* NBER Working Paper No. 24877.

Lux T (1995). *Herd behavior, bubbles and crashes,* Economic Journal, 105:881‐896.

Lux T (1998). *The socio-economic dynamics of speculative markets: Interacting agents, chaos, and the fat tails of*

*return distributions,* Journal of Economic Behavior and Organization, 33:143-165.

Lux T, Marchesi M (1999). *Scaling and criticality in a stochastic multi-agent model of a financial market,* Nature, 397:498 – 500.

Lux T, Marchesi M (2000). *Volatility clustering in financial markets: A micro- simulation of interacting agents,* International Journal of Theoretical and Applied Finance, 3:67-702.

Kim G, Markowitz HM (1989). *Investment rules, margin and market volatility,* Journal of Portfolio Management. 16:45-52.

Makarov I, Schoar A (2020). *Trading and arbitrage in cryptocurrency markets,* Journal of Financial Economics, 135, 2, 293-319.

Mandelbrot B (1966). *Nonlinear forecasts, rational bubbles, and martingales,* Journal of Business 39, 242-255.

Perepelitsa (2021). *Psychological dimension of adaptive trading in cryptocurrency markets,* arXiv:2109.12166.

Perepelitsa M and Timofeyev I (2019). *Asynchronous stochastic price pump,* Physica A: J. Stat. Mech. Appl. 516, 356-364.

Perepelitsa M and Timofeyev I (2021). *Self-sustained price bubbles driven by digital currency innovations and adaptive market behavior,* SN Business and Finance.

Roth AE, Erev I (1995). *Learning in extensive-form games: experimental data and simple dynamics models in the intermediate term,* Games and Economic Behavior, 8 164-212.

Scheinkman JA, Wei X (2003). *Overconfidence and speculative bubbles*, J. Political Economy 111, 1183-1220.

Sha W, Vergne JP (2017). *Buzz factor or innovation potential: What explains cryptocurrencies returns?* PloS One 12 (1):e0169556.

Sornette D (2003). *Why Stock Markets Crash: Critical Events in Complex Financial Systems.* Princeton University Press.

Stigler GJ (1964). *Public regulation of the securities market,* Journal of Business, 37.

Tedeschi G, Iori G, Gallegati M (2012). *Herding effects in order driven markets: The rise and fall of gurus,* J. Econ. Behav. Org., vol. 81, no. 1, pp. 82-96.